# ESTRATEGIA DE RECUPERACIÓN DE DATOS DESDE EL ÚLTIMO NIVEL REDUCIENDO EL TIEMPO DE REGISTRO EN EL SOFTWARE

## BOTTOM-UP STRATEGY FOR DATA RETRIEVAL AND DATA ENTRY OVER FRONT-END APPLICATION SOFTWARE

**Rusel Cierto Trinidad** [1]

**Alcides Bernardo Tello** [2]

Universidad Nacional Hermilio Valdizan – Perú

**ABSTRACT**

Some people implement "pattern" and "best practices" without analysing its efficiency on their projects. Consequently, our goal in this article is to convince software developers that it is worth to make an earnest effort to evaluate the use of best practices and software patterns. For such purpose, in this study we took a concrete case system for geographical locations inputs through user interfaces. Then, we performed a comparative study on a traditional method against our approach, named "reverse logistic" to retrieve results, by measuring the time that a user spends to perform actions when entering data into a system. Surprisingly, we had a decrease of 59% in the amount of time spent in comparison to the time spent on the traditional method. This result lays a foundation for feeding data from the typical final step and search based on string matching algorithms, speeding up the interaction between people and computer response.

**RESUMEN**

Algunas personas implementan "modelos" y "mejores prácticas" sin analizar su eficiencia en sus proyectos. En consecuencia, nuestro objetivo en este artículo es convencer a los desarrolladores de software que vale la pena hacer un esfuerzo serio para evaluar el uso de las mejores prácticas y modelos de software.

---

1     Rusel Cierto Trinidad Universidad Nacional Hermilio Valdizan – Perú
rciertot@outlook.com.pe /rcierto@unheval.edu.pe ,

2     Alcides Bernardo Tello Universidad Nacional Hermilio Valdizan – Perú
a.btello@hotmail.com / abernardo@unheval.edu.pe

Para tal fin, en este estudio tomamos un sistema de casos concretos para entradas de ubicaciones geográficas a través de interfaces de usuario. Luego, realizamos un estudio comparativo sobre un método tradicional con nuestro enfoque, denominado "logística inversa" para recuperar resultados, al medir el tiempo que un usuario dedica a realizar acciones al ingresar datos en un sistema. Sorprendentemente, tuvimos una disminución del 59% en la cantidad de tiempo empleado en comparación con el tiempo empleado en el método tradicional. Este resultado establece una base de datos para la alimentación de la etapa final típica y búsqueda basada en algoritmos de coincidencia de cadenas, la aceleración de la interacción entre las personas y la respuesta del ordenador.

**Palabras clave**: Recuperación de datos, interfaz de usuario, interacción hombre-máquina, interfaz, capa de acceso, diseño de interacción, entrada de datos.

**INTRODUCTION**

The primary goals of user interface design is to create user interfaces which are user-friendly, self-explanatory and efficient, [John14] and [Tidw10]. In our study we have focused on efficiency with respect to time spent during data entry. Data entry allude to user tasks involving both input of data to a computer and computer responses to those inputs.

Inefficiencies engendered by poorly designed data entry transactions are so visible, many users interface design in clerical jobs deal with data entry enquiries [Rand16].

The greatest need, however, in current information systems is for enhancing the logic of data entry [Fend15]. Thus, the study presented here deal with data entry algorithms, insofar as possible, in the absence of attention to their hardware implementation

Data can be entered into a computer in a wide assortment of ways. The easiest type of data entry consists merely of pointing at something. In more complex kinds of data entry is the control of the format of data inputs by the user. In general, in the industrial design [Rand94] and in human–computer interaction, the aspiration is to allow effective operation. Human–Computer Interaction is the study of the way in which computer technology influences human work and activities [DIX 09].

For effective operation and efficiency, in this study, we selected several websites that use the coding system for geographical locations, called Ubigeo.

The remaining section in this paper are organized as follow:

In section 2, we state some assumption to be true in this article.

In section 3, we explore the case: Traditional method or top down approach for data retrieval. Then we also measure the time consumed by the user when selecting data to input into a system. Examining the nature of each website in relation to the data retrieval, the processes through which we acquire data from servers by using queries, we found a common top-down feature for choosing administrative subdivision: regions, provinces, districts. To test, we have developed a bottom-up approach as it is described in section 4.

Section 5, we discuss the time consumption measured in section 4 and 3, in order to ensure comparability.

Finally, we draw conclusion in the section 6.

## 2 ASSUMPTIONS

In this section we assumed the following propositions to be true statement for this study.

With the advent of robustness and reliability of computer power, more work is carried out by the front-end activities in web applications [Tidw10].

The only way a person interacts with a computer, tablet, smartphone or other electronic device is trough user interface [Loud10].

The blind assumption that introducing new techniques must be good.

Algorithms for recovering data from structured database that resembles google search engines are easily implemented [Serg14] and [Greh14]

## 3 TOP-DOWN APPROACH FOR DATA RETRIEVAL: TRADITIONAL METHOD

The first level of geographical location database starts with country, followed by administrative subdivisions: regions, provinces, districts as second, third and fourth level respectively.

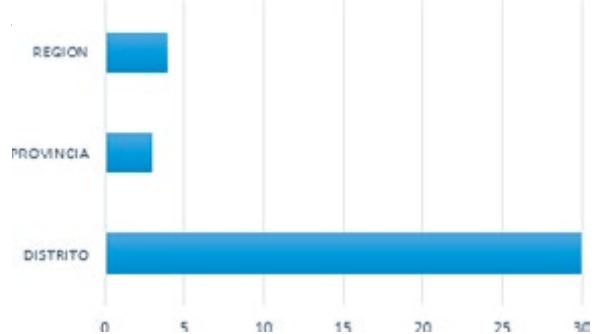

Figure 01: Traditional data entry for geographical locations

Figure 01 shows typical data entry related to identifiable natural person. Lets focus on Date of birth (lugar de Nacimiento) at located at fifth row. It begins with country with a dropdown combo, then region (Departamento), province (provincia) and district (distrito). A dropdown combo enables the user to select either by typing text into it or by selecting an item from a displayed list. Obviously, for each level of selection, there is a query that fetches a subset of place names from the database and populates the next level combobox afterwards using such data retrieval.

Table 01: Time spent for each level of geographic location using traditional method.

| Region | 30 milliseconds |
|---|---|
| Provincia | 40 milliseconds |
| Dsitrito | 21 milliseconds |
| TOTAL | 91 milliseconds |

We have measured the number of milliseconds elapsed for each level of selection as shown in table 01. As it is shown we need on average 91 milliseconds for our complete search. A bar graph is also provided below in the Figure 02.

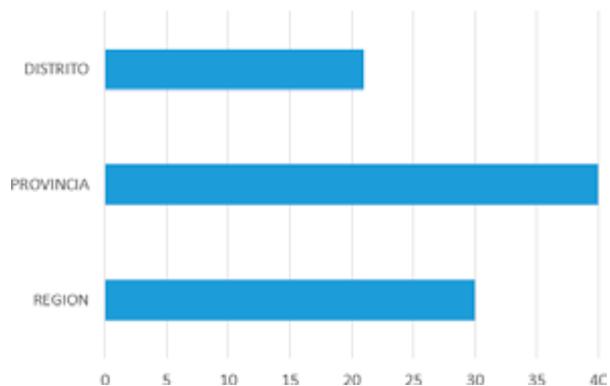

Figure 02: Time spent in traditional data entry

## 4 PROPOSED METHOD: REVERSE LOGISTIC FOR DATA RETRIEVAL

In this section we present a new way of thinking when developing software for data retrieval: Begin backwards from the root.

We advocate for a bottom up approach when retrieving data from geographical location database. - e.g. typing the root of the location. We only need any graphical control element intended to enable the user to input text information, e.g. a text box, text field or text entry box, dropdown combo. They are always available in every programming language and web design forms.

Figure 03: Reverse logistic for data retrieval during data entry for geographical locations

The Figure03 shows the implementation of our approach. We start with the lowest level entry, in this case district (distrito). The program displays immediate results below the edit area as we type the text. As we start to type our search, it anticipates what we are looking for and begins to show results for our search. Then, once we choose the item, the remaining level of search will immediately be populated with the correct data.

Moreover, the underlying algorithm looks only for clues to give us back a list of choices from which we decide exactly what we want to select, for this search is based on string matching algorithms.

Table 02: Time spent for each level of geographic location using reverse logistic method.

| Region | 30 milliseconds |
|---|---|
| Provincia | 3 milliseconds |
| Dsitrito | 4 milliseconds |
| TOTAL | 37 milliseconds |

We have also measured the number of milliseconds for this method and it is shown in table 02. The total time is

37 milliseconds. The Figure 04 illustrates the same information using a bar graph.

Figure 04: Time spent reverse logistic for data retrieval during data entry for geographical locations

In section 3, inefficiencies arise from running a query processes in each level of geographic location, yielding four separate sets of location.

## 5 DRAWING A COMPARISON BETWEEN TRADITIONAL DATA RETRIEVAL AND REVERSE LOGISTIC AND ITS JUSTIFICATION.

In this section, we scrutinize the results from the previous two sections and make clear the time-saving result.

First, from the Table01 and Table02, we find the actual amount of the decrease in the time spent in each case, which is 54 milliseconds; because 91 milliseconds are reduced to 37.

This number corresponds to 59% of the total time spent in the traditional data entry and retrieval method.

$$(0.59)(91) \cong 54$$

This clearly informs us that less time is spent in the front end when a user's input data by utilizing our reverse regression approach. We lowered the time by 59%.

In addition, when entering data into the system by using our logistic approach, the information appears instantly rather than slowly descending from the top of the hierarchical geographic locations.

For many years, we have been implementing geographical location search in the top-down fashion almost without thinking because it has become a habit, pattern or best practice sometime in the past. It has justification in early computational capabilities when computer was far slower than recent versions.

From this perspective, patterns and best practices can become obsolete or irrelevant after a time. It happens so faster with the increasingly rapid pace of technological development towards the end of the twentieth century and our present century. To make matters worse, developer had not learnt how to deal with these rapid changes in their undergraduate programs.

We are confident with this result since it has a justification in Aristotle's Law of Identity.

As in the standard terminology in database theory, each element is called entity. Hence, each geographical location is an entity. Therefore, it has a justification in Aristotle's Law of Identity where each entity exists as something and peculiar or specific. In our proposed approach we begin with the entity at the lowest level to recover information for the highest level.

## 6 CONCLUSIONS

By changing the logic of data entry algorithm, we have reduced the time spent by the user when feeding in data into a database system.

We have revealed how patterns and best practices in software engineering can become obsolete in time due to the faster improvement in bandwidth and programming language performance. Therefore, it is worth an earnest effort to evaluate the use of best practices and software patterns on their projects before implementing them, accordingly with the advent of increasingly powerful computer. It allows us a new method or practice, undergoing a change from what we are doing now to doing something different, it can even be the reverse manner.

We only focused on time efficiency aspect of Human- Computer Interaction in this study. The other aspects such as user-friendly, self-explanatory requires behavioural sciences, design, media studies, and so forth; which is beyond the purview of four knowledge. Knowing the mentioned area, may move us to examine the efficiency and flexibility further.

**BIBLIOGRAPHIC REFERENCES**